\documentclass{emulateapj}

\makeatletter

\begin{document}
\title{Dynamic Alignment and Exact Scaling Laws in Magnetohydrodynamic Turbulence}
\author{Stanislav Boldyrev}
\affil{Department of Physics, University of Wisconsin at Madison, 1150 University Ave, 
Madison, WI 53706; {\sf boldyrev@wisc.edu}}
\author{Joanne Mason} 
\author{Fausto Cattaneo}
\affil{Department of Astronomy and Astrophysics, University of Chicago, 
5640 S. Ellis Ave, Chicago, IL 60637; {\sf jmason@flash.uchicago.edu; cattaneo@flash.uchicago.edu}}
\date{\today}

\begin{abstract}
Magnetohydrodynamic (MHD) turbulence is pervasive in astrophysical systems. Recent high-resolution numerical simulations 
suggest that the energy 
spectrum of strong incompressible MHD turbulence is $E(k_{\perp})\propto k_{\perp}^{-3/2}$.  
So far, there has been no phenomenological theory that simultaneously explains this spectrum and satisfies the exact analytic 
relations for MHD turbulence due to Politano \& Pouquet. Indeed, the Politano-Pouquet relations are often invoked to suggest 
that the spectrum of MHD turbulence instead has the Kolmogorov scaling~$-5/3$.
Using geometrical arguments and numerical tests, here we analyze this seeming contradiction and demonstrate that the $-3/2$ 
scaling and the Politano-Pouquet relations are reconciled by the phenomenon of scale-dependent dynamic alignment that was 
recently discovered in MHD turbulence.

\end{abstract}


\maketitle

\section{Introduction}
Magnetized plasma turbulence plays an essential role in many astrophysical phenomena,  
such as the solar wind \citep[e.g.][]{goldstein}, interstellar scintillation \citep[e.g.][]{lithwick}, 
cosmic ray acceleration, propagation and scattering 
in the interstellar medium \citep[e.g.,][]{kulsrud,wentzel} and thermal conduction in galaxy 
clusters \citep[e.g.,][]{rechester,chandran,narayan}. 
The statistical properties of such turbulence can be inferred either indirectly from astronomical observations, such as 
scintillation of interstellar radio sources, or from in situ measurements, such as measurements of the magnetic and velocity 
fields in the solar wind. 
Over a wide range of scales, turbulence in astrophysical plasmas can be modeled in the framework of incompressible 
magnetohydrodynamics. When written in
terms of the Els\"asser variables the equations have the form 
\begin{eqnarray}
 & \left(\frac{\partial}{\partial t}+{\bf v}_A\cdot\nabla \right){\bf 
  z}+\left({\bf w}\cdot\nabla\right){\bf z} = -\nabla P, \label{mhd1} \\
 &  \left(\frac{\partial}{\partial t}-{\bf v}_A\cdot\nabla \right){\bf 
  w}+\left({\bf z}\cdot\nabla\right){\bf w} = -\nabla P,
\label{mhd2}
\end{eqnarray}
where the Els\"asser variables are defined as ${\bf z}={\bf 
v}-{\bf b}$ and ${\bf w}={\bf v}+{\bf b}$,  ${\bf v}$ is the fluctuating plasma velocity, ${\bf b}$ is
the fluctuating magnetic field normalized by $\sqrt{4 \pi \rho_0}$,
${\bf v}_A={\bf B}_0/\sqrt{4\pi \rho_0}$ is the Alfv\'en velocity corresponding to the uniform magnetic field ${\bf B}_0$, 
the ``total'' pressure $P=p/\rho_0+b^2/2$ includes the 
plasma pressure $p$ and the magnetic pressure and $\rho_0$ is the
background plasma density that we assume to be constant. 

Current theoretical understanding of MHD turbulence largely relies on phenomenological models and numerical simulations 
\citep[e.g.,][]{biskamp}. However, there are certain exact results in the statistical theory of turbulence that can be used to test 
these predictions. The results have attracted considerable interest recently, due to a number of solar wind observations where 
the exact relations were directly tested \citep[e.g.,][]{vasquez07,macbride08,marino08,podesta09}. In what follows we discuss 
the exact relations formulated for MHD turbulence, the so-called Politano-Pouquet relations, and we analyze to what extent 
they agree with the recent phenomenological and numerical predictions for the energy spectrum of strong anisotropic MHD 
turbulence.      

It is well known that the Kolmogorov theory for isotropic incompressible hydrodynamic turbulence yields the following exact 
relation for the third-order longitudinal structure function of the velocity field in the inertial range \citep[see, e.g.,][]{frisch}:
\begin{eqnarray}
\langle \delta v_L^3({\bf r})  \rangle = -\frac{4}{5} \epsilon r.
\label{kolmogorov}
\end{eqnarray}
Here $\delta {v}_L({\bf r})=[{\bf v}({\bf x}+{\bf r})-{\bf v}({\bf x})]\cdot {\bf r}/r$ is the longitudinal component of the 
velocity difference between two points separated by the vector ${\bf r}$ and $\epsilon$ is the rate of energy supply to 
the system at large scales. In a stationary state it coincides with 
the rate of energy cascade toward small dissipative scales and with 
the rate of energy dissipation. If one assumes that the fluctuations are not 
strong compared to the rms value of $\delta {\bf v}({\bf r})$, and that $\delta v(r)\sim \delta v_L(r)$, 
one can dimensionally estimate 
from (\ref{kolmogorov}) that $\langle \delta v^2({\bf r})\rangle \propto r^{2/3}$. The Fourier transform of the 
latter expression 
then leads to the Kolmogorov spectrum for the turbulent velocity field, $E(k)\propto k^{-5/3}$. 

Interestingly, analogous relations hold for isotropic magnetohydrodynamic turbulence. \cite{politano,politano2} derived 
\begin{eqnarray}
S^w_{3L}(r)\equiv \langle \delta z_L (\delta {\bf w})^2\rangle=-\frac{4}{3}\epsilon^{w}r,
\label{pp1}\\
S^z_{3L}(r)\equiv \langle \delta w_L (\delta {\bf z})^2\rangle=-\frac{4}{3}\epsilon^{z}r,
\label{pp2} 
\end{eqnarray}
where $\delta z_L$ and $\delta w_L$ are the longitudinal components of $\delta {\bf z}$ and $\delta {\bf w}$, $\epsilon^w$ is 
the transfer rate of the ${\bf w}$ field and $\epsilon^z$ is the transfer rate of the ${\bf z}$ field. If one now follows the 
analogy with the hydrodynamic case and 
assumes that all typical fluctuations scale in the same way ($\delta z_L \propto \delta w_L \propto  \delta z \propto \delta w 
\propto \delta v \propto \delta b$) one derives  
$\delta v_r\propto \delta b_r \propto r^{1/3}$, which leads to the Kolmogorov 
scaling of the MHD turbulence spectrum. 

The results (\ref{pp1},\ref{pp2}) can be extended to the case of anisotropic turbulence in the presence of a strong guiding 
field -- a setting relevant for astrophysical turbulence where a large-scale field is always present, whether due to external 
sources or large-scale eddies \citep[see, e.g.,][]{maron,matthaeus,milano}.  According to \cite{politano}, the requirement of 
homogeneity allows one to derive the following differential relations in the inertial range of turbulence
\begin{eqnarray}
\frac{\partial}{\partial r^i} \langle \delta z^i (\delta {\bf w})^2\rangle= -4 \epsilon^w ,
\label{pp3}\\
\frac{\partial}{\partial r^i} \langle \delta w^i (\delta {\bf z})^2\rangle= -4 \epsilon^z.
\label{pp4} 
\end{eqnarray}
Expressions (\ref{pp1},\ref{pp2}) immediately follow if the correlation functions are assumed to be three-dimensionally 
isotropic. However, in the case with a strong guiding field the variations of the fluctuations in the field perpendicular direction 
are much stronger than their field parallel variations. Hence the latter can be neglected in the inertial interval and the spatial 
derivatives in (\ref{pp3}, \ref{pp4}) can be replaced by their field perpendicular parts
\begin{eqnarray}
\frac{\partial}{\partial r_{\perp}^i} \langle \delta z^i (\delta {\bf w})^2\rangle= -4 \epsilon^w ,
\label{pp5}\\
\frac{\partial}{\partial r_{\perp}^i} \langle \delta w^i (\delta {\bf z})^2\rangle= -4 \epsilon^z,
\label{pp6} 
\end{eqnarray}
where ${\bf r}_{\perp}$ is a vector in the field-perpendicular plane and the variations of the fields $\bf z$ and $\bf w$ are 
taken along ${\bf r}_{\perp}$, for example $\delta {\bf w}\equiv {\bf w}({\bf x}+{\bf r}_{\perp})-{\bf w}({\bf x})$. 
Assuming statistical isotropy in the field-perpendicular plane one then derives, analogously to 
(\ref{pp1},\ref{pp2})
\begin{eqnarray}
\langle \delta z_{L}(\delta {\bf w})^2\rangle = -2\epsilon^w r_{\perp}, 
\label{pp7}\\
\langle \delta w_{L}(\delta {\bf z})^2\rangle = -2\epsilon^z r_{\perp}.
\label{pp8}
\end{eqnarray}
Here the longitudinal components of ${\bf z}$ and ${\bf w}$ are defined along the two-dimensional vector ${\bf r}_{\perp}$, 
for example $\delta z_L \equiv [{\bf z}({\bf x}+{\bf r}_{\perp})-{\bf z}({\bf x})]\cdot {\bf r}_{\perp}/r_{\perp}$.  
A more formal derivation of expressions analogous to~(\ref{pp7},\ref{pp8}) can be found in \cite{perez-boldyrev}.

Arguments similar to those described for the hydrodynamic case may lead one to conclude that the energy spectrum for MHD 
turbulence with a strong guide field is also $E(k_{\perp})\propto k_{\perp}^{-5/3}$ \citep[see, e.g., the discussion and 
references in][]{biskamp,verma}. Phenomenological arguments leading to such a spectrum have attracted considerable 
attention \citep[][]{higdon,goldreich,verma}. This however reveals a puzzling contradiction with recent high resolution 
numerical simulations of strongly magnetized turbulence that instead suggest $E(k_{\perp})\propto 
k_{\perp}^{-3/2}$ \citep[see][]{maron,biskamp-muller,muller,mason,mason2}. Although phenomenological arguments 
explaining such a spectrum have been proposed \citep{boldyrev,boldyrev2} so far it remained unclear whether they can be 
reconciled with the exact Politano-Pouquet relations. This apparent inconsistency motivated our interest in the problem. 

The goal of the present paper is to analyze whether the field-perpendicular energy spectrum $E(k_{\perp})\propto 
k_{\perp}^{-3/2}$ is consistent with the exact Politano-Pouquet relations. We propose that the spectral 
exponent ${-3/2}$ does not, in fact, contradict the rigorous Politano-Pouquet result. Rather, their  
relation is manifested in the phenomenon of scale-dependent dynamic alignment that was recently discovered in driven MHD 
turbulence.

\section{Dynamic alignment and Politano-Pouquet relations}
\label{alignment}
Dynamic alignment is a known phenomenon of MHD 
turbulence~\citep[e.g.,][]{dobrowolny,grappin,pouquet,pouquetsm,politanops}.  
However, in previous studies it essentially meant that decaying MHD turbulence asymptotically reaches the so-called Alfv\'enic 
state where either ${\bf v}({\bf x})\equiv {\bf b}({\bf x})$ 
or ${\bf v}({\bf x})\equiv -{\bf b}({\bf x})$, depending on initial conditions. It has been realized 
recently that the effect is modified in randomly driven MHD turbulence, where the fluctuations $\delta {\bf v}_r$ and 
$\pm\delta {\bf b}_r$ tend 
to align their directions in such a way that the alignment angle becomes {\em 
scale-dependent}~\citep{boldyrev,boldyrev2,mason}. 

The essence of the phenomenon is that at each  
field-perpendicular scale $r$ ($\sim 1/k_{\perp}$) in the inertial range, typical 
shear-Alfv\'en velocity fluctuations ($\delta {\bf v}_{r}$) and magnetic fluctuations ($\pm \delta {\bf b}_{r}$) tend 
to align the directions of their polarizations in the 
field-perpendicular plane and the turbulent eddies become anisotropic in that plane. In such eddies the  magnetic and velocity 
fluctuations change significantly in the direction almost perpendicular to the directions of the fluctuations themselves,~$\delta 
{\bf v}_r$ and~$\pm \delta {\bf b}_r$, which reduces the strength of the nonlinear interaction in the MHD equations. A 
numerical illustration of this phenomenon can be found in \cite{perez-boldyrev2}. The alignment and anisotropy  
are stronger for smaller scales, with the alignment angle decreasing with scale 
as $\theta_{r}\propto r^{1/4}$. This leads to the velocity and magnetic fluctuations $\delta v_{r}\sim \delta b_r\propto 
r^{1/4}$ and hence the energy spectrum $E(k_{\perp})\propto k_{\perp}^{-3/2}$, discussed in the introduction. We will now 
show its effect on the Politano-Pouquet relations.

There are two possibilities for the dynamic alignment:  
the velocity fluctuation $\delta {\bf v}_r$ can be aligned either 
with $\delta {\bf b}_r$ (positive alignment) or with $-\delta {\bf b}_r$ (negative alignment). This implies that the turbulent 
domain is fragmented into regions of positive and negative alignment. If no overall alignment 
is present, the numbers of positively and negatively aligned eddies are balanced on average. In the case of nonzero overall 
alignment there is an imbalance of positively and negatively aligned eddies. 

However, as found in \cite{perez-boldyrev2}, there is no essential difference between overall balanced and imbalanced strong 
turbulence. Therein it is argued that strong MHD turbulence, whether overall balanced or not, has the characteristic property 
that at each scale it is locally imbalanced. 
Overall, it can be viewed as a superposition of positively and negatively aligned eddies.  The scaling of the turbulent energy 
spectrum depends on the way the alignment changes with scale, not on the amount of overall alignment. 
 
Let us now determine which of the two configurations (positive or negative alignment) provides the dominant contribution to 
the structure 
functions~(\ref{pp1}) and (\ref{pp2}).  We note that, by definition, the amplitudes of $\delta {\bf v}_r$ and $\delta {\bf b}_r$ 
are of the order of their typical, rms values. In the case when $\delta {\bf v}_r$ is aligned with $\delta {\bf b}_r$, we have 
$\delta w_r > \delta z_r$. Therefore, this configuration contributes more to structure function~(\ref{pp1}) than to structure 
function~(\ref{pp2}). Similarly, the configuration in which  $\delta {\bf v}_r$ is aligned with $-\delta {\bf b}_r$, and hence  
$\delta z_r > \delta w_r$, provides the dominant contribution to structure 
function~(\ref{pp2}).

\begin{figure} [tbp]
\begin{center}
\includegraphics[scale=0.5]{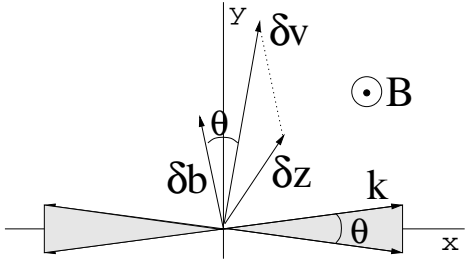}
\end{center}
\vskip-1mm
\caption{Sketch of the velocity and the magnetic field fluctuations, $\delta {\bf v}_{r}$ and $\delta {\bf b}_r$,  aligned in the 
field-perpendicular plane within a small angle~$\theta_r$.} 
\vskip2mm
\label{align_fig_1}
\end{figure}

Without loss of generality we consider in detail only 
the structure function $S^w_{3L}(r)$, 
defined in~(\ref{pp1}), and we concentrate on the contribution provided by 
the configuration in which $\delta {\bf v}_r$ is aligned with $\delta {\bf b}_r$ inside a turbulent eddy. Figure~\ref{align_fig_1} 
illustrates this case. 
The large-scale field ${\bf B}_0$ is in the z-direction and the vectors $\delta {\bf v}_{r}$ 
and $\delta {\bf b}_{r}$ are aligned within a small angle $\theta_{r}$ in the 
field-perpendicular plane, in the y-direction, say. 
Since the polarization of shear-Alfv\'en waves are perpendicular to $\mathbf{B_0}$ and to their wave vector $\mathbf{k}$, 
the wave vectors ${\bf k}$ are aligned 
in the x-direction here. Consequently, the variation 
of the fields is strongest in the x-direction and the dominant contribution to the structure 
function~(\ref{pp1}) comes from the situation in which 
the point-separation 
vector ${\bf r}$  lies in the x-direction. 
It follows from geometrical arguments that the longitudinal projection (i.e., x-component) of $\delta {\bf z}_r$, $\delta z_L$, is 
smaller than the typical value of $\delta w_r$ by a factor of order $\theta_{r}$, {\it viz.} $\delta w_r \sim \delta v_r$ and 
$\delta z_L \sim \theta_r \delta v_r$. This introduces an extra factor of $\theta_{r}$ in Politano-Pouquet correlation 
function~(\ref{pp1}), and one obtains
\begin{eqnarray} 
\langle \delta z_L (\delta {\bf w})^2\rangle 
\sim \theta_r \delta v^3_{r}. 
\label{scaling}
\end{eqnarray}

As was demonstrated in \cite{boldyrev,boldyrev2} and \cite{mason,mason2},  
the {\em scale-dependent} dynamic 
alignment $\theta_{r}\propto r^{1/4}$ leads to the scaling of the fluctuating 
fields $\delta v_{r}\sim \delta b_{r}\propto r^{1/4}$, which 
explains the numerically observed field-perpendicular energy 
spectrum $E(k_{\perp})\propto k_{\perp}^{-3/2}$.  
Quite remarkably, by substituting these 
scalings into expression~(\ref{scaling}) we also satisfy the scaling relation~(\ref{pp1}). 
Thus, the numerical findings are reconciled  
with the Politano-Pouquet relations if one invokes the phenomenon 
of scale-dependent dynamic alignment. 
 
In the next section we attempt to test relation~(\ref{scaling}) using numerical simulations of strong incompressible MHD 
turbulence.
In particular, we measure the following third-order structure functions 
\begin{eqnarray}
& {\tilde S}_{3L}^w(r)=\langle |\delta z_L| (\delta {\bf w})^2\rangle,
\label{S3L}\\
& S_3^w(r)=\langle|\delta {\bf w}|^3\rangle.
\label{S3}
\end{eqnarray}
We use the absolute value of $\delta z_L$ in calculating~(\ref{S3L}) to avoid cancellations and slow convergence caused by 
different signs of $\delta z_L$.   
If our idea expressed by~(\ref{scaling}) is correct then the functions ${\tilde S}^w_{3L}(r)$ and $S_3(r)$ should have 
essentially {\em different} scalings, and, as follows from the estimate ${\tilde S}^w_{3L}(r)\sim \theta_r \delta v_r^3$ and  
$S_3(r)\sim \delta v_r^3$, 
their ratio should correspond to the alignment angle~$\theta_r$, and therefore should scale as
\begin{eqnarray}
{\tilde S}^w_{3L}(r)/S^w_{3}(r)\propto r^{0.25}.
\label{ratio}
\end{eqnarray}
Note that the scaling of the alignment angle $\theta(r)\propto r^{0.25}$ can be measured independently with the aid of  
second-order structure functions \citep[e.g.,][]{mason,mason2,beresnyak-lazarian}. \\

\section{Numerical results}
We solve the incompressible MHD equations (\ref{mhd1}, \ref{mhd2})
using standard pseudospectral methods. 
An external magnetic field is applied in $z$-direction with strength $B_0\approx 5$ measured in units of velocity. The periodic 
domain has a resolution of $512^3$ mesh points and is elongated in the $z$-direction with aspect ratio 1:1:$B_0$.  An external 
force ${\bf f}({\bf x}, t)$, and small fluid viscosity $\nu$ and resistivity $\eta$ are added to the equations. The external force 
is random and it drives the turbulence at large scales. The details of the numerical method and set up can be found in 
\cite{mason2}. 

The Reynolds number is defined as $Re=U_{rms}L/\nu$,
where $L$ $(\sim 1)$ is the field-perpendicular box size, $\nu$ is fluid viscosity
and $U_{rms}$ $(\sim 1)$ is the rms
value of velocity fluctuations.
We restrict ourselves to the case in which the magnetic resistivity and fluid viscosity are the same, $\nu=\eta$, with $Re\approx 
2200$. The system is evolved until a statistically steady state is reached, which is confirmed by observing the time evolution of 
the total energy
of the fluctuations. The data set consists of 30 samples that cover approximately 6 large-scale eddy turnover times. 
\begin{figure} [tbp]
\hskip-0.5cm
\begin{center}
\includegraphics[scale=0.35]{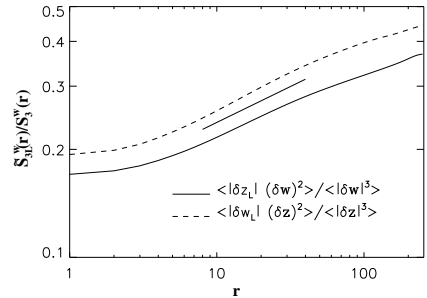}
\end{center}
\vskip-5mm
\caption{The relative scaling of the structure functions ${\tilde S}^w_{3L}(r)$ and $S^w_3(r)$. The ratio (\ref{ratio}) is 
plotted versus scale~$r$. 
The equivalent procedure with ${\bf w}$ and ${\bf z}$ interchanged yields a similar  
slope. 
The straight line has the slope~$0.2$.}
\vskip1mm
\label{align_scaling_fig}
\end{figure}
To calculate the structure functions~(\ref{S3L}) and~(\ref{S3}) we 
construct $\delta {\bf z}(r)={\bf z}({\bf x}+{\bf r})-{\bf z}({\bf x})$ 
and $\delta {\bf w}(r)={\bf w}({\bf x}+{\bf r})-{\bf w}({\bf x})$,  where ${\bf r}$ is in a plane perpendicular 
to~${\bf B}_0$. By definition, $\delta z_{L}=\delta {\bf z}(r)\cdot {\bf r}/r$ 
and $\delta w_{L}=\delta {\bf w}(r)\cdot {\bf r}/r$. The average is taken over different positions of 
the point ${\bf x}$ in that plane, over all such planes in the data cube, and then over all data cubes.

The results we present here correspond to one of five simulations described in \cite{mason2} (Case 2a). Each of those 
simulations differ by the large-scale driving mechanism, which is not expected to affect the turbulent dynamics in the inertial 
interval. The numerical calculation 
of expression~(\ref{ratio}) is shown in Figure~\ref{align_scaling_fig}. For this case we find ${\tilde 
S}^w_{3L}(r)/S^w_{3}(r)\propto r^{0.2}$. 
Repeating the calculation for the other four cases yields noticeable scatter from case to case, with 
the slopes ranging from approximately $0.17$ to $0.21$. We discuss the results in the next section. 

\begin{figure} [tbp]
\begin{center}
\includegraphics[scale=0.4]{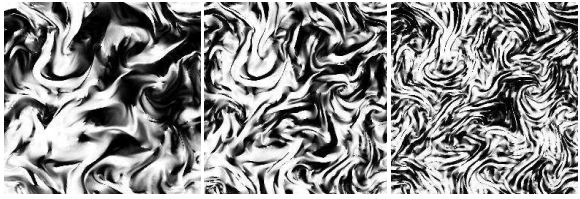}
\end{center}
\caption{Alignment regions at different scales. Plotted is $\cos(\theta)=(\mathbf{\tilde v} \cdot \mathbf{\tilde 
b})/|\mathbf{\tilde v}| |\mathbf{\tilde b}|$ in a plane perpendicular to the guiding field $\mathbf{B_0}$, where $\mathbf{\tilde 
v}$ and $\mathbf{\tilde b}$ are the velocity and magnetic field filtered at the scale $k_f$. Left--right: $k_f=1,4,10$.}
\label{alignment_regions}
\end{figure}

\section{Discussion and conclusion}
We have proposed phenomenological arguments demonstrating that the energy spectrum $E(k_{\perp})\propto 
k_{\perp}^{-3/2}$ for MHD turbulence is consistent with the exact analytic results that are known as the 
Politano-Pouquet relations. 
We argue that the two results are reconciled by the phenomenon of scale-dependent dynamic alignment. As a numerical 
illustration, we computed the structure functions (\ref{S3L}) and~(\ref{S3}). We found that their scalings are essentially 
different, which supports our arguments presented in section~\ref{alignment}. However, the agreement with the analytic 
prediction ${\tilde S}^w_{3L}(r)/S^w_{3}(r)\propto r^{0.25}$ turns out to be only qualitative. Several reasons 
may explain the lack of good quantitative agreement. First, this may be due to extremely slow convergence of the statistics for 
the third-order structure functions in (\ref{ratio}). 
Indeed we noticed that in each run the convergence of the statistics for the measured third-order structure functions is quite 
slow. The evolution of the slopes from snapshot to snapshot apparently has long-time variations. 
In spite of the large statistical ensemble accumulated in our runs, it is not enough for the precise measurement 
of the slope in~(\ref{ratio}). 
Second, the size of the Reynolds number in our simulations (which is limited by computational costs) may significantly impede 
computation of the slope. Third, there may be a systematic deviation of the slope measured in~(\ref{ratio}) due to 
intermittency corrections to the third-order structure function~(\ref{S3}), which are not captured by our model.  We plan to 
address these issues in future numerical work. 

Finally, in Figure~\ref{alignment_regions} we illustrate the regions of alignment at different scales. It appears that majority of 
the domain is covered with regions of high positive or negative alignment, and that this structure is hierarchical --  within a 
large-scale region of positive alignment, say, there exist smaller scale highly aligned and anti-aligned structures. This supports 
our assumption that even overall balanced MHD turbulence is imbalanced locally, so that different spatial regions are 
responsible for different structure functions in the Politano-Pouquet relations. We note that the presence of correlated 
polarized regions in MHD turbulence was also observed in some early simulations~\citep[e.g.,][]{maron}.   

\acknowledgments
This work was supported by the NSF Center for Magnetic
Self-Organization in Laboratory and Astrophysical Plasmas
at the University of Wisconsin at Madison and the University of Chicago. The work of SB is supported by the Department of 
Energy under the Grant No. DE-FG02-07ER54932. This research used resources of the Argonne Leadership Computing 
Facility at Argonne National Laboratory, which is supported by the Office of Science of the U.S. Department of Energy under 
contract DE-AC02-06CH11357.

\end {document}